\documentclass[prc,aps,twocolumn,
amsmath,amssymb,amsfonts,widetable] {revtex4-1}
\usepackage{bm}
\usepackage{soul}
\usepackage{array}
\usepackage{lipsum}
\usepackage{relsize}
\usepackage{mathrsfs}
\usepackage{amssymb}
\usepackage{amsmath}
\usepackage{graphicx}
\usepackage[usenames]{color}
\usepackage{pslatex}

\def\Cevens{CE$\nu$NS\hspace{3pt}}

\newcommand{\Lagrange}[1]{\mathscr{L}_{\raisebox{-0.5pt}{\scriptsize{#1}}}}

\begin{document}
\title{From CREX to CE$\nu$NS: The Weak Radius of ${}^{40}$Ar} 
\author{J. Piekarewicz}\email{jpiekarewicz@fsu.edu}
\affiliation{Department of Physics, Florida State University,
               Tallahassee, FL 32306, USA}
\date{\today}
\begin{abstract}
 Despite significant theoretical efforts, the CREX-PREX dilemma remains unresolved,
 preventing the reliable prediction of neutron (or weak-charge) radii that, besides their 
 intrinsic nuclear-structure interest, often serve to quantify the impact of nuclear uncertainties 
 in searches for new physics. Coherent elastic neutrino-nucleus scattering is a clean 
 and attractive portal to new physics whose sensitivity may be impacted by such nuclear 
 uncertainties. In this paper we use CREX as our main anchor, together with a strong 
 calcium-argon correlation, to provide a robust baseline for the weak radius of 
 ${}^{40}$Ar: $R_{\rm wk}^{\,40} = 3.452 \pm 0.028~\text{(stat)} \pm 0.022~\text{(syst)}\,\text{fm}$.
 The weak radius of argon is an observable highly relevant to ongoing and future liquid-argon 
 campaigns that encodes the loss of coherence at small momentum transfers.
 \end{abstract}

\smallskip
\pacs{
21.10.Gv, 
21.60.Jz,       
25.30.Bf,       
}
\maketitle

\section{Introduction}
\label{sec:Introduction}

Shortly after the discovery of weak neutral currents in 1973, coherent elastic neutrino–nucleus 
scattering (\Cevens\!) was proposed as a process with favorable cross sections that could 
illuminate a variety of astrophysical phenomena, such as neutrino transport in core-collapse 
supernovae and neutron stars\,\cite{Freedman:1973yd}. \Cevens is considered “favorable” 
because its cross section scales with the square of the weak charge of the nucleus, which is 
dominated by the neutron number. Consequently, \Cevens has emerged as a powerful probe 
of the weak-charge density. In principle, \Cevens can provide information on the spatial distribution 
of neutrons in a nucleus in much the same way that elastic electron scattering has mapped
the proton distribution. In practice, however, such a task is considerably more challenging.

Despite being proposed by Freedman in 1973, \Cevens eluded experimental confirmation for 
more than four decades. The breakthrough came at the Spallation Neutron Source of Oak Ridge 
National Laboratory, where a high-intensity neutrino beam was directed onto a low-background 
CsI scintillator\,\cite{Akimov:2017ade}. The long delay---despite the coherent enhancement of the 
cross section---was due to the enormous experimental challenge of detecting the low-energy 
nuclear recoil. Unlike in electron scattering, the scattered lepton in \Cevens can not be observed 
directly, making the detection of the recoiling nucleus the only observable signature.

Although enormously challenging from an experimental perspective, the \Cevens cross section 
is remarkably simple. Besides a few kinematical factors, the cross section for a spinless target 
depends solely on the Fermi constant, the weak charge of the nucleus, and its associated weak 
form factor; see Eq.(\ref{CEvens1}). Because of this simplicity, \Cevens finds broad applications 
across nuclear structure\,\cite{Patton:2012jr,Patton:2013nwa,Cadeddu:2017etk,Yang:2019pbx,
Hoferichter:2020osn,Cadeddu:2023tkp}, fundamental symmetries\cite{AristizabalSierra:2018eqm,AristizabalSierra:2019zmy,Papoulias:2019xaw,
Abdullah:2022zue}, dark matter searches\cite{Hoferichter:2015ipa,Aalbers:2022dzr,
XENON:2024ijk}, and supernova detection\,\cite{Horowitz:2003cz,Scholberg:2012id,Ko:2022pmu}, 
among others\,\cite{Akimov:2017ade}. Given \Cevens wide range of applications, it is imperative to 
quantify nuclear-structure uncertainties associated with the weak form factor of the target nucleus. 
Motivated by the recent progress of the COHERENT Collaboration in measuring for the first time 
\Cevens on argon\,\cite{Akimov:2019rhz,COHERENT:2020ybo,COHERENT:2020iec}, we devote 
the present work on constraining the weak nuclear radius of ${}^{40}$Ar\,($R^{\,40}_{\rm wk}$). 
Note that at the small momentum transfers probed in these experiments, the mild deviations from 
full coherence are imprinted on the weak radius of the target nucleus. That is, 
\begin{equation}
 F_{\rm wk}(Q^{2})\!=\!1-\frac{1}{6}Q^{2}R_{\rm wk}^{2}+\ldots
 \label{FF}
\end{equation}

The main strategy employed in this work is to use a set of accurately calibrated covariant energy 
density functionals\,\cite{Fattoyev:2013yaa,Chen:2014sca,Chen:2014mza,Salinas:2023nci} to 
predict the weak-charge density of ${}^{40}$Ar, the associated form factor and ultimately the weak 
radius. Most of these models are accompanied by a covariance matrix, enabling the estimation of 
statistical averages, uncertainties, and correlations. However, significant differences exist among 
these models---particularly in the isovector sector---due to the scarcity of nuclear observables with 
large neutron-proton asymmetries. As a result, the models span a relatively broad range of values 
for the slope of the symmetry energy, a quantity that strongly impacts the neutron 
skin\,\cite{Brown:2000,Furnstahl:2001un,RocaMaza:2011pm}, and consequently, the weak radius 
of neutron-rich nuclei. 

Recently, however, significant progress has been made in constraining the density dependence of 
the symmetry energy. Building on a 30-year-old proposal by Donnelly, Dubach, and 
Sick\,\cite{Donnelly:1989qs}, two experimental campaigns at the Thomas Jefferson National Accelerator 
Facility successfully measured the weak form factor of both $^{208}$Pb\,\cite{Abrahamyan:2012gp,
Horowitz:2012tj,Adhikari:2021phr} and $^{48}$Ca\,\cite{Adhikari:2022kgg} at appropriate values of the 
momentum transfer. These experiments used parity-violating electron scattering, a purely electroweak 
reaction that is free from the many uncertainties that affect hadronic probes. The PREX collaboration 
reported the following value for the weak radius of $^{208}$Pb\,\cite{Adhikari:2021phr}:
\begin{equation}
R_{\rm wk}^{208}\!=\!(5.800 \pm 0.075)\,\text{fm}.
\label{PREX}
\end{equation}
Leveraging the well-established correlation between the neutron skin thickness of $^{208}$Pb 
($R_{\rm skin}^{208}$) and the slope of the symmetry energy at saturation density 
($L$)\,\cite{Brown:2000,Furnstahl:2001un,RocaMaza:2011pm}, a relatively large value of 
$L\!=\!(106 \pm 37)\,\text{MeV}$ was inferred\,\cite{Reed:2021nqk}. This implies that the equation 
of state of pure neutron matter near saturation density is stiff, thereby suggesting that low-mass 
neutron stars should have relatively large radii\,\cite{Horowitz:2001ya,Carriere:2002bx}.

Under the assumption that the slope of the symmetry energy also controls the neutron skin 
thickness of ${}^{48}$Ca, one would have anticipated a fairly large neutron skin in calcium. 
Instead---and to the surprise of many---the CREX Collaboration reported a significantly smaller 
``weak skin'' defined as the difference between the weak and charge radii\,\cite{Adhikari:2022kgg}. 
Once combined with its known charge radius\,\cite{Angeli:2013}, the following value for the weak 
radius of ${}^{48}$Ca is obtained:
\begin{equation}
   R_{\rm wk}^{\,48} = (3.636 \pm 0.035)\,\text{fm}.
   \label{CREX}
\end{equation}

\begin{figure}[ht]
\smallskip
 \includegraphics[width=0.85\columnwidth]{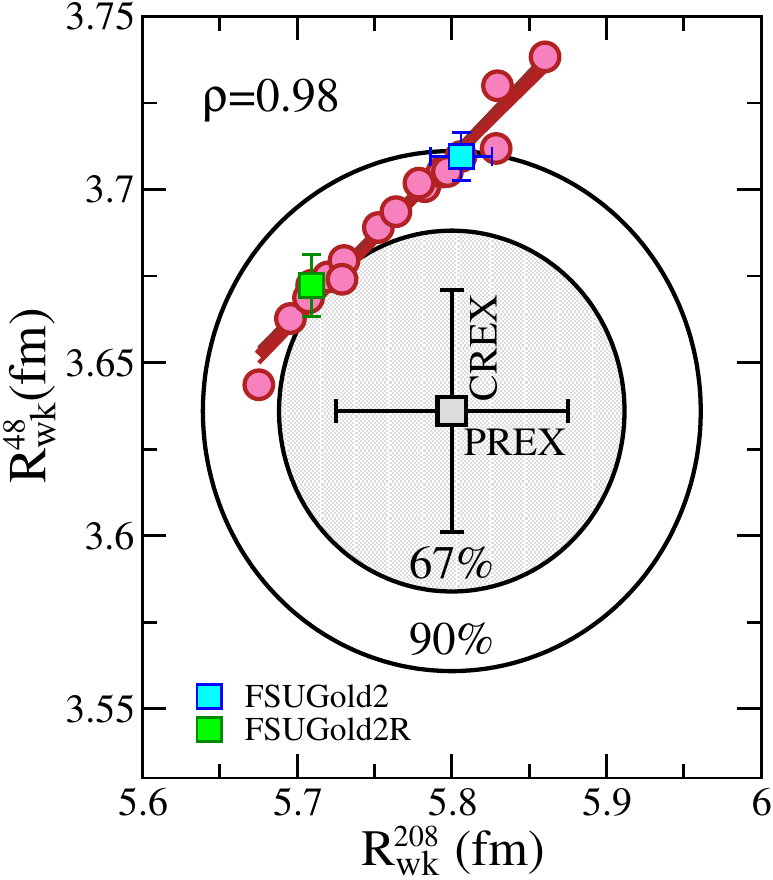}
\caption{(Color online) Predictions for the weak radius of ${}^{208}$Pb and ${}^{48}$Ca
               from the 17 covariant energy density functionals used in this work. The 
               ellipses represent joint PREX and CREX 67\% and 90\% probability contours, 
               respectively. The FSUGold2 functional\,\cite{Chen:2014sca} and its refinement
               FSUGold2R\,\cite{Salinas:2023nci} illustrate typical statistical uncertainties.}
\label{Fig1}
\end{figure}

To highlight the ``CREX–PREX dilemma,'' we display in Fig.\ref{Fig1} the experimentally determined 
weak radii of $^{48}$Ca and $^{208}$Pb as ellipses representing the 67\% and 90\% confidence regions. 
Also shown are predictions from 17 covariant energy density functionals that will be employed in this 
work. As anticipated, the theoretical correlation between the two weak radii is strong ($\rho\!=\!0.98$): 
models that predict a large radius in $^{208}$Pb, as suggested by PREX, invariably predict a large 
radius for $^{48}$Ca, contradicting the CREX result. At best, only a few models graze the 67\% confidence 
ellipse. The situation is even more dramatic when considering the difference between the charge and 
weak form factors; see Fig.\,2 in Ref.\,\cite{Adhikari:2022kgg}.

Despite enormous efforts by the theoretical community\,\cite{Hu:2021trw,Reinhard:2022inh,
Zhang:2022bni,Mondal:2022cva,Papakonstantinou:2022gkt,Yuksel:2022umn,Li:2022okx,
Thakur:2022dxb,Miyatsu:2023lki,Reed:2023cap,Sammarruca:2023mxp,Salinas:2023qic,
Yue:2024srj,Zhao:2024gjz,Roca-Maza:2025vnr,Kunjipurayil:2025xss}, no compelling 
resolution to the CREX–PREX dilemma has yet emerged. Consequently, one may be led 
to believe that predicting the weak radius of $^{40}$Ar from existing theoretical models is 
doomed to failure. To circumvent this difficulty, we adopt a different strategy that relies 
heavily on the CREX result. Specifically, we will demonstrate the existence of a robust 
correlation between the weak radii of the two medium-mass nuclei, ${}^{48}$Ca and 
${}^{40}$Ar. Relying primarily on this correlation and little else, the weak radius of 
${}^{40}$Ar will be directly inferred from the CREX measurement.

The paper is organized as follows. In Sec.\,\ref{sec:Formalism} we provide a brief discussion 
of the physics underlying \Cevens, emphasizing its role as a model-independent probe of 
neutron densities and as a potential portal to new physics. This section also summarizes the 
main elements of the theoretical framework adopted in the present work. In Sec.\,\ref{sec:Results} 
we present results obtained within the covariant density functional theory framework. Finally, 
Sec.\,\ref{sec:Conclusions} offers a summary of the main findings together with perspectives 
for future work.


\section{Formalism}
\label{sec:Formalism}

\subsection{Coherent elastic neutrino-nucleus scattering}
\label{sec:CEvNS}

\begin{figure}[ht]
\smallskip
 \includegraphics[width=0.7\columnwidth]{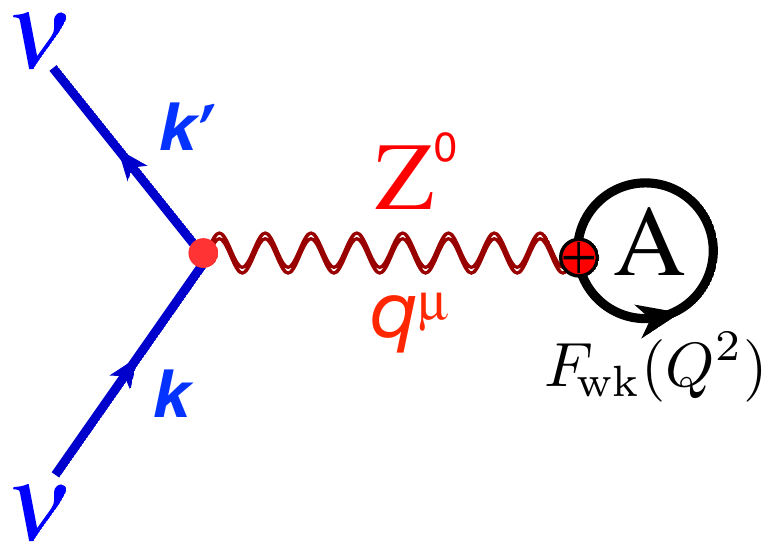}
\caption{(Color online) Feynman diagram for the elastic scattering of neutrinos from a spinless 
              nuclear target. Information on the internal structure of the nucleus is entirely contained 
              in the weak nuclear form factor $F_{\rm wk}(Q^{2})$.} 
\label{Fig2}
\end{figure}

In contrast to the long and successful tradition of elastic electron scattering as a sensitive probe 
of  proton densities, electroweak processes have only recently been used to map the neutron 
distribution\,\cite{Abrahamyan:2012gp,Horowitz:2012tj,Akimov:2017ade,Adhikari:2021phr,
Adhikari:2022kgg}. Because the weak charge of the proton is small, both parity-violating electron 
scattering (PVES) and \Cevens serve as ideal probes of neutron densities. Yet the extremely weak 
nature of the interaction makes such experiments enormously challenging. In the particular case of 
\Cevens illustrated in Fig.\,\ref{Fig2}, the only viable path to full kinematical reconstruction is the 
detection of nuclear recoils of extremely low energy. 

Although the experimental measurement is challenging, the standard-model derivation of the \Cevens
cross section is relatively simple. Following previous work\,\cite{Yang:2019pbx}, the ground-state 
matrix element of the weak neutral current can be written as
\begin{equation}
  \langle p' \vert J_{\rm NC}^{\mu} \vert p \rangle
  = Q_{\rm wk}\,F_{\rm wk}(Q^{2})\,(p+p')^{\mu},
  \label{ChFF}
\end{equation}
where $p$ ($p'$) is the initial (final) four-momentum of the target nucleus, $Q^{2}$ is the square of the 
four-momentum transfer, $Q_{\rm wk}\!=\!-N\!+\!(1\!-\!4\sin^{2}\!\theta_{\rm W})Z$ is the weak nuclear 
charge, and the form factor has been normalized to one at zero momentum transfer. By contracting the 
leptonic and hadronic tensors, one obtains the differential cross section in the laboratory frame in terms 
of the kinetic energy $T$ of the recoiling nucleus\,\cite{Scholberg:2005qs}:
\begin{equation}
  \left(\frac{d\sigma}{dT}\right)
  = \frac{G_{F}^{2}}{8\pi}\,M
    \left[ 2 - 2\,\frac{T}{E} - \frac{M T}{E^{2}} \right]
    Q_{\rm wk}^{2}\,F_{\rm wk}^{2}(Q^{2}),
  \label{CEvens1}
\end{equation}
where $G_{\!F}$ is the Fermi constant, $M$ is the mass of the target nucleus, $E$ is the incident neutrino 
energy, and $Q^{2}\!=\!2MT$. Note that the differential cross section at forward angles (small $Q^{2}$) 
is \emph{coherent}, namely, the amplitudes from all nucleons add constructively, leading to a cross section 
that scales approximately with the \emph{square} of the number of neutrons. The loss of coherence, namely,
the suppression of the cross section at larger momentum transfers, is encoded in the weak nuclear form 
factor $F_{\rm wk}(Q^{2})$. 

\subsection{Covariant density functional theory}
\label{sec:CovDFT}

All the results presented in this paper are computed within the framework of covariant density functional 
theory (DFT). In this approach, and for the specific version adopted here, the underlying Lagrangian density 
may be expressed as
\begin{equation}
\Lagrange{} = \Lagrange{0} + \Lagrange{1} + \Lagrange{2},
\end{equation}
where $\Lagrange{0}$ denotes the non-interacting component, consisting of the kinetic energy of all the
constituents (nucleons and mesons).

The second term, $\Lagrange{1}$, contains the Yukawa couplings between nucleons and mesons. These 
nucleon-meson couplings are formulated in terms of scalar and vector bilinears of both isoscalar and isovector 
character that couple to the corresponding meson fields to preserve Lorentz covariance. That is,
\begin{equation}
\Lagrange{1} =
\bar\psi \left[g_{\rm s}\phi   - 
    \left(g_{\rm v}V_\mu  \!+\!
    \frac{g_{\rho}}{2}{\mbox{\boldmath $\tau$}}\cdot{\bf b}_{\mu} \!+\!    
    \frac{e}{2}(1\!+\!\tau_{3})A_{\mu}\right)\!\gamma^{\,\mu}
         \right]\psi 
 \label{L1}
\end{equation}
where $\psi$ denotes the isodoublet nucleon field and $\boldsymbol{\tau}$ is the vector containing the three 
Pauli matrices, with $\tau_{3}$ being its $z$-component. The short-range nuclear interaction is mediated by 
two isoscalar mesons: a scalar field ($\sigma$), which provides the intermediate-range attraction, and a vector 
field ($V_{\mu}$), which generates the short-range repulsion. The isovector-vector field (${\bf b}_{\mu}$) 
accounts for the isospin dependence of the nuclear force, while the photon field ($A_{\mu}$) mediates the 
long-range Coulomb repulsion. 

Finally, $\Lagrange{2}$ incorporates meson self-interactions that simulate density-dependent 
effects. This contribution contains both unmixed and mixed meson nonlinearities that have been systematically 
introduced over time\,\cite{Boguta:1977xi,Serot:1984ey,Mueller:1996pm,Lalazissis:1996rd,Serot:1997xg,
Horowitz:2000xj,Todd-Rutel:2005fa,Chen:2014sca,Chen:2014mza} to refine the original Walecka 
model\,\cite{Walecka:1974qa}. That is, $\Lagrange{2}$ is given by
\begin{widetext}
\begin{equation}
\Lagrange{2} =    - \frac{1}{3!} \kappa\,\Phi^3 - \frac{1}{4!} \lambda\Phi^4 
                            + \frac{1}{4!} \zeta (W_\mu W^\mu)^2 
                            + \Lambda_{\rm v} (\boldsymbol{B}_\mu \cdot \boldsymbol{B}^{\,\mu})(W_\mu W^\mu),
\end{equation}
\end{widetext}
where $\Phi\!\equiv\!g_{\rm s}\phi$, $W_\mu\!\equiv\!g_{\rm v}V_\mu$, and 
$\boldsymbol{B}_\mu\!\equiv\!g_{\rho}\boldsymbol{b}_\mu $. A detailed discussion of the physical role of each 
of these terms may be found in Ref.\,\cite{Salinas:2023qic} and references contained therein.

To assess the impact of the CREX and PREX campaigns on the weak radius of ${}^{40}$Ar, we employ 
a set of 17 covariant energy density functionals (EDFs)\,\cite{Reed:2021nqk,Salinas:2023nci}. The main 
differences among these models reside in the isovector sector, which controls the density dependence of 
the nuclear symmetry energy. In particular, this set of functionals spans values of the slope of the symmetry 
energy from $L\!=\!47\,\text{MeV}$ to $L\!=\!135\,\text{MeV}$, corresponding to a range in the neutron-skin 
thickness of  ${}^{208}$Pb of $R_{\rm skin}^{208}\!=\!(0.15\!-\!0.33)\,\text{fm}$. 

In the following section, we 
use these 17 EDFs to quantify systematic theoretical uncertainties. Such uncertainties reflect the intrinsic 
limitations of the theoretical framework, arising from the specific choice of functional, missing physics, or 
the lack of experimental data to further constrain the model. These should be distinguished from genuine 
statistical uncertainties, which originate from the finite precision with which a given set of model parameters 
are calibrated to existing experimental data. In this paper we address both statistical and systematic
uncertainties.

From the set of 17 covariant EDFs, we single out two---FSUGold2\,\cite{Chen:2014sca} and 
FSUGold2R\,\cite{Salinas:2023nci}---to carry out a statistical analysis of correlations. The calibration  of 
FSUGold2 relied on three classes of observables: (i) ground-state properties of magic and semi-magic 
nuclei, (ii) centroid energies of giant monopole resonances, and (iii) the maximum mass of neutron stars. 
By incorporating information on monopole energies, but none on neutron skin thicknesses, FSUGold2 
predicted a soft equation of state for symmetric nuclear matter alongside a relatively stiff symmetry energy. 

Motivated by the remarkable progress achieved in recent years, we refined FSUGold2 by incorporating 
additional constraints from three complementary sources: (i) tidal deformabilities extracted by the LIGO-Virgo 
collaboration\,\cite{Abbott:PRL2017,Abbott:2018exr}, (ii) stellar radii inferred by the NICER 
mission\,\cite{Riley:2019yda,Miller:2019cac,Miller:2021qha,Riley:2021pdl}, and (iii) the equation of state 
of pure neutron matter as predicted by chiral effective field theory (EFT)\,\cite{Drischler:2021kxf}. The refined 
functional, dubbed FSUGold2R\,\cite{Salinas:2023nci}, is particularly sensitive to the chiral-EFT input, as the 
relatively large uncertainties in the astrophysical data limit their impact. Indeed, chiral EFT drives most of the 
refinement, as both the symmetry energy at saturation and its slope are sharpened and reduced. Results for 
various bulk properties of infinite nuclear matter as predicted by FSUGold2 and its refinement FSUGold2R 
are listed on Table\,\ref{Table1}.

\begin{center}
\begin{table}[h]
\begin{tabular}{|l|c|c|c|c|c|}
\hline\rule{0pt}{2.5ex}   
  \!Model & $\mathlarger{\mathlarger{\rho}}_0$ & $\mathlarger{\mathlarger{\varepsilon}}_{0}$ & $K$ & $J$ & $L$  \\
    \hline\hline
    FSUGold2  & 0.151(1) & -16.28(2) & 238.0(28) & 37.62(111) & 112.8(161) \\
    FSUGold2R& 0.152(1) & -16.22(2) & 241.2(25) & 32.03(23)\;\;  & 57.2(10)  \\
    \hline \hline
\end{tabular}
\caption{Bulk properties of infinite nuclear matter as predicted by two accurately calibrated models: 
              FSUGold2\,\cite{Chen:2014sca} and its refinement FSUGold2R\,\cite{Salinas:2023nci}. 
              The bulk properties listed are the baryon density, energy per nucleon, compressibility, 
              symmetry energy, and slope of the symmetry energy---all evaluated at saturation density. 
              All quantities are given in MeV except the density that is quoted in $\text{fm}^{-3}$.}
\label{Table1}
\end{table}
\end{center}
 
Because the neutron excess is modest for most of the ground-state observables used to calibrate 
FSUGold2, the bulk properties of symmetric nuclear matter ($\mathlarger{\rho}_0$, 
$\mathlarger{\varepsilon}_{0}$, and $K$) remain practically unchanged. By contrast, the isovector 
sector undergoes a dramatic change: both the symmetry energy and its slope at saturation density 
become significantly reduced and much more tightly constrained. In particular, the slope 
parameter---so critical in determining neutron skins and stellar radii of low mass stars---decreases 
from $L\!=\!(112.8 \pm 16.1)\,\text{MeV}$ to $L\!=\!(57.2 \pm 1.0)\,\text{MeV}$. As we show in
the next section, the refined interaction---without solving the CREX-PREX dilemma---finds a compromise 
between the large neutron skin thickness of ${}^{208}$Pb and the much smaller skin of ${}^{48}$Ca. 

\section{Results}
\label{sec:Results}

\subsection{Statistical Uncertainties}
\label{sec:Statistical}

We start this section by focusing on statistical uncertainties and correlations encoded in the covariance 
matrices obtained from the calibration of FSUGold2\,\cite{Chen:2014sca} and its subsequent refinement 
FSUGold2R\,\cite{Salinas:2023nci}. Both covariance matrices are available from the author upon
request. 

\begin{figure}[ht]
\smallskip
 \includegraphics[width=0.95\columnwidth]{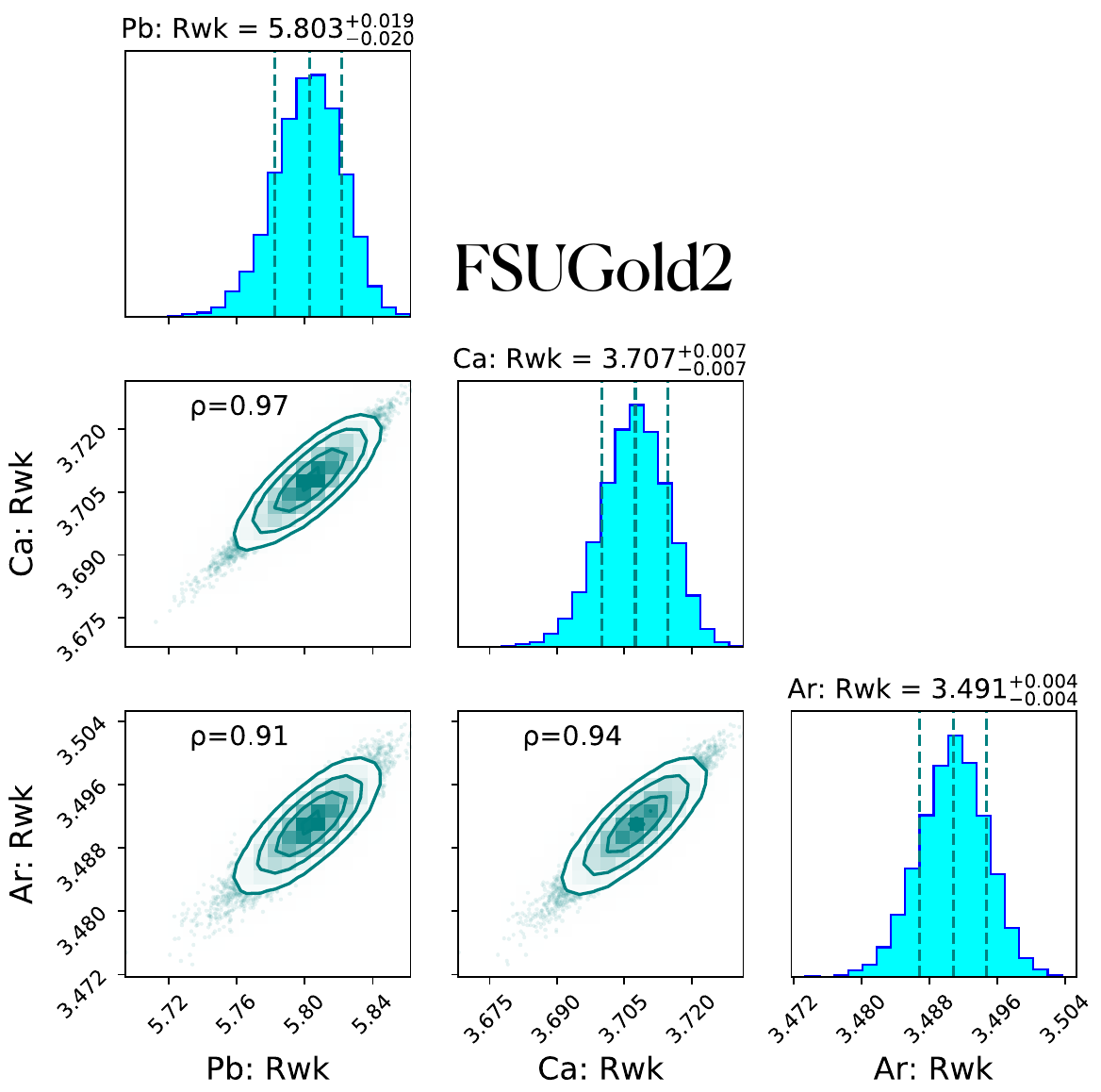}
\caption{(Color online) Corner plot of weak radii from the covariance matrix associated to the FSUGold2 
                                     EDF\,\cite{Chen:2014sca}. The diagonal panels show one-dimensional
                                     distributions for the weak radii of ${}^{208}$Pb ${}^{48}$Ca, and ${}^{40}$Ar,
                                     respectively. The off-diagonal panels display the corresponding joint posteriors,
                                     with the innermost contour corresponding to the 1$\sigma$ region, and the numbers 
                                     on the panels denoting the correlation coefficients.}
\label{Fig3}
\end{figure}

In Fig.\ref{Fig3} averages and correlation coefficients are displayed for the weak radii of ${}^{208}$Pb, 
${}^{48}$Ca, and ${}^{40}$Ar.  Because no strong isovector indicator was incorporated into the calibration 
of the functional, FSUGold2 predicts a stiff symmetry energy, consistent with earlier parametrizations such 
as NL3\,\cite{Lalazissis:1996rd}. Consequently, the prediction for the weak radius of ${}^{208}$Pb, namely,
$R_{\rm wk}^{\,208}\!=\!(5.803 \pm 0.020)\,\text{fm}$, is in complete agreement with the PREX result listed 
in Eq.\,(\ref{PREX}) and from which a stiff symmetry energy was inferred\,\cite{Reed:2021nqk}. In contrast, 
the corresponding prediction for ${}^{48}$Ca of $R_{\rm wk}^{\,48}\!=\!(3.707 \pm 0.007)\,\text{fm}$, lies 
well outside the $1\sigma$ range of the CREX measurement; see Eq.\,(\ref{CREX}) and Fig.\,\ref{Fig1}.
Given the strong correlation displayed in Fig.\ref{Fig3} between calcium and argon, it is highly likely that 
the predicted value for the yet-to-be-measured weak radius of ${}^{40}$Ar will also be overestimated.

\begin{figure}[ht]
\smallskip
 \includegraphics[width=0.95\columnwidth]{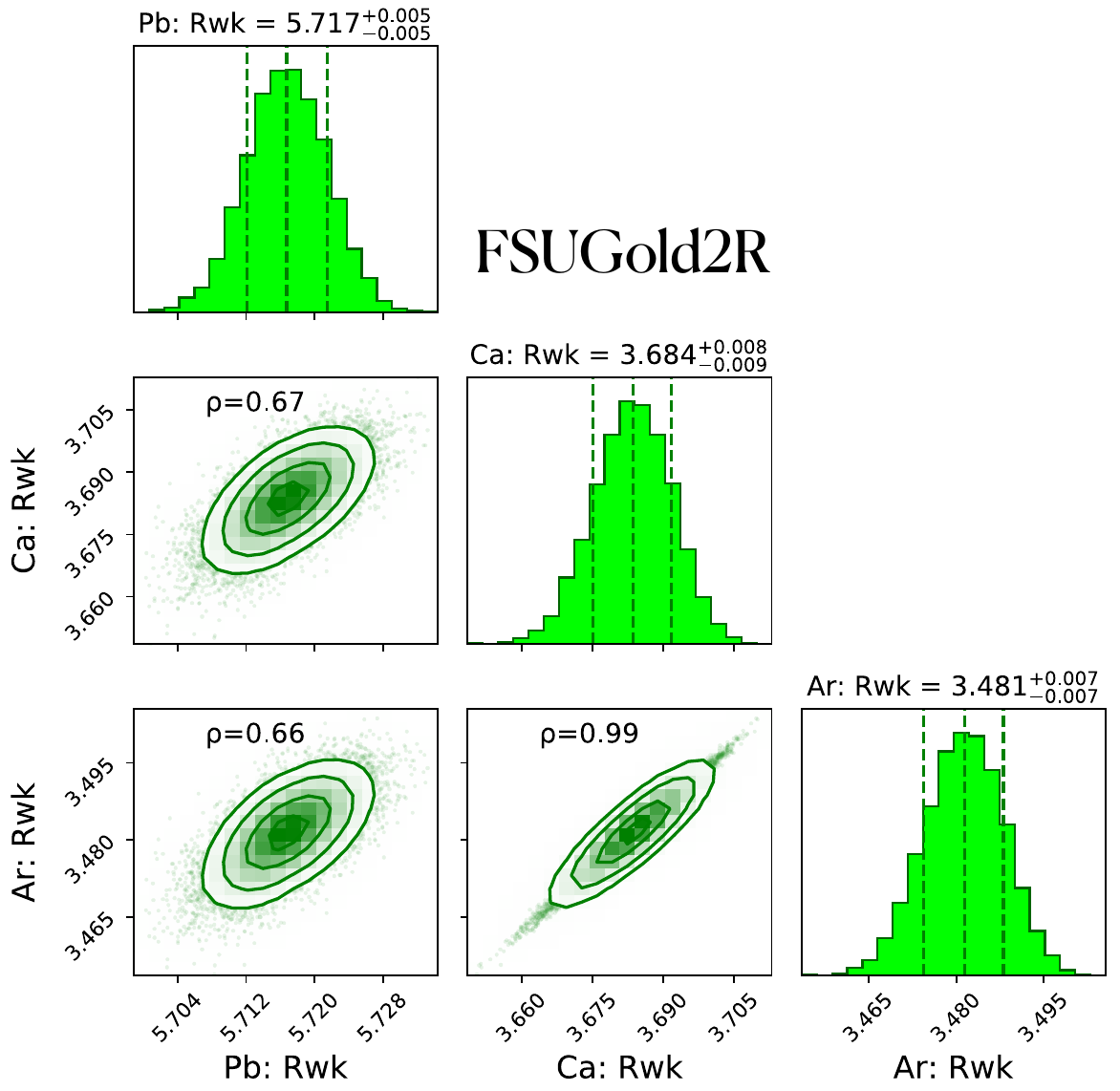}
\caption{(Color online) Corner plot of weak radii from the covariance matrix associated to the FSUGold2R 
                                     EDF\,\cite{Salinas:2023nci}. The diagonal panels show one-dimensional
                                     distributions for the weak radii of ${}^{208}$Pb ${}^{48}$Ca, and ${}^{40}$Ar,
                                     respectively. The off-diagonal panels display the corresponding joint posteriors,
                                     with the innermost contour corresponding to the 1$\sigma$ region, and the numbers 
                                     on the panels denoting the correlation coefficients.}
\label{Fig4}
\end{figure}

In Fig.\ref{Fig4}, we display the same information as in Fig.\ref{Fig3}, but now for the refined FSUGold2R 
functional. The strong correlation between the weak radii of lead and calcium seen earlier is now significantly 
weakened. Although this may appear to offer a potential resolution to the CREX-PREX dilemma---given that 
the FSUGold2R predictions now lie within the 67\% confidence interval (see Fig.\ref{Fig1})---the more likely 
interpretation is that these results represent a compromise between PREX and CREX, rather than a genuine 
resolution of the conflict. Nevertheless, in the next section we will continue to exploit the almost perfect correlation 
between calcium and argon predicted by FSUGold2R.

\vspace{1in}

\subsection{CREX Informing \Cevens}
\label{sec:CIC}
The previous sections illustrated that covariant energy density functionals---at least of the kind considered 
here---are unlikely to provide a reliable prediction for the weak radius of argon, a fundamental quantity that 
encodes the first departure from full coherence in the \Cevens cross ection. Fortunately, there is no need to 
rely exclusively on theory. By combining the CREX result with the strong correlation displayed by FSUGold2R 
in Fig.\ref{Fig4}, one can provide a robust prediction for the weak radius of ${}^{40}$Ar.

To illustrate the procedure, we first consider the ideal case in which the correlation between the two weak radii is 
perfect; we will adjust later for small deviations from $\rho\!\equiv\!1$. Assuming the CREX result is normally
distributed, the induced distribution for the weak radius of argon is also normal with mean and standard deviation
\begin{equation}
  \mu_{{}_{40}} = a + b\,\mu_{{}_{48}} \hspace{5pt}\text{and}\hspace{5pt} 
  \sigma_{{}_{40}} = b\,\sigma_{{}_{48}},
 \label{LinCorr1}
\end{equation}
where $\mu_{{}_{48}}$ and $\sigma_{{}_{48}}$ are the CREX values in Eq.\,(\ref{CREX}), and $a$ and $b$ are the 
intercept and slope obtained from the (assumed) perfect linear relation.

In the realistic case that the correlation is not perfect, the mean $\mu_{{}_{40}}$ remains unchanged, while the 
variance must be corrected to account for uncertainties in $a$ and $b$, their covariance, and any residual scatter.
That is,
\begin{equation}
  \sigma_{{}_{40}}^{2}\!=\!\sigma_{a}^{2}\!+\!(\mu_{48}^{2}+\sigma_{48}^{2})\sigma_{b}^{2}
                            + b^{2}\sigma_{48}^{2} +2\,\mu_{{}_{48}}\mathrm{Cov}(a,b)+\sigma_{\rm int}^{2},
 \label{LinCorr2}
\end{equation}
where $\sigma_{a}^{2}$, $\sigma_{b}^{2}$, and $\mathrm{Cov}(a,b)$ are the elements of the $2\!\times\!2$ 
linear-regression covariance matrix, and $\sigma_{\rm int}^{2}$ is the intrinsic (residual) variance. 

\begin{figure}[ht]
\smallskip
 \includegraphics[width=0.7\columnwidth]{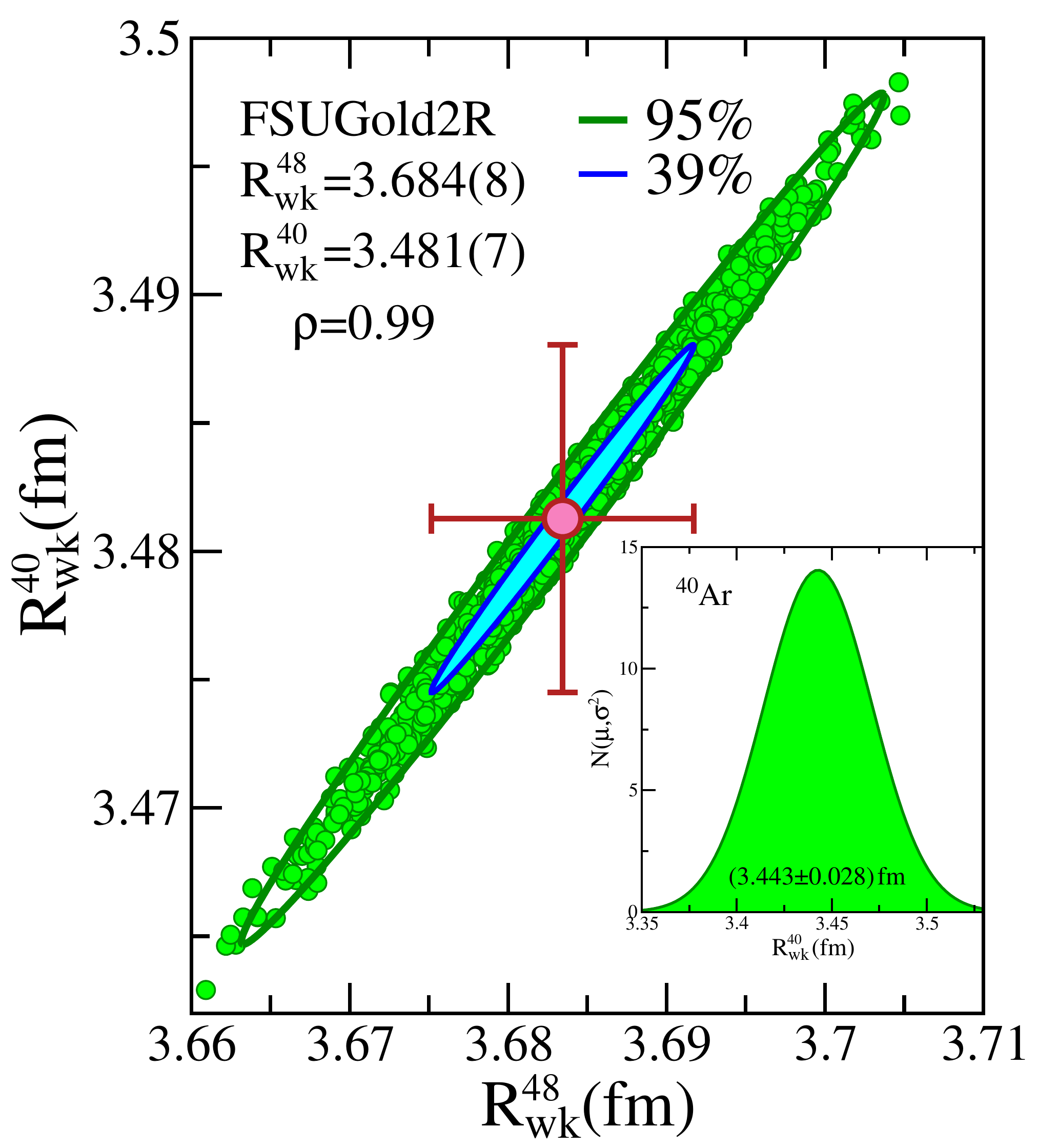}
\caption{(Color online) Joint posterior for the weak radii of ${}^{48}$Ca and ${}^{40}$Ar obtained from
               10,000 MCMC samples drawn from the multivariate normal distribution obtained from the 
               FSUGold2R calibration\,\cite{Salinas:2023nci}. Contours enclose 39\% (the $1\sigma$ 
               region) and 95\% of the joint probability. The inset shows the CREX-informed normal distribution 
               for $R_{\rm wk}^{\,40}$; see text for details.}
\label{Fig5}	                         
\end{figure}

These results are summarized in Fig.\ref{Fig5}, which displays the 39\% and 95\% confidence 
ellipses, with the 39\% contour corresponding to the $1\sigma$ region for a bivariate normal
distribution. Also shown with $1\sigma$ error bars, are the mean values and standard deviations 
predicted by the FSUGold2R functional. As already established, the theoretical prediction for 
$R_{\rm wk}^{\,48}$ overestimates the reported experimental value; see Fig.\ref{Fig1}. Hence, 
rather than using the predicted value directly, we reduce model dependence by invoking only the 
strong correlation between the two weak radii. Using Eqs.(\ref{LinCorr1})-(\ref{LinCorr2}) results
in the normal distribution shown in the inset of Fig.\ref{Fig5}, which yields a CREX-informed estimate 
for the weak radius of ${}^{40}$Ar:
\begin{equation}
 R_{\rm wk}^{\,40} = (3.443 \pm 0.028)\,\text{fm}.
 \label{Rwk40Stat}
\end{equation}
Because the correlation coefficient is nearly one, the elements of the linear-regression covariance matrix 
are small, so the quoted uncertainty is dominated by the product $b\,\sigma_{{}_{48}}$, as in the ideal 
case of Eq.(\ref{LinCorr1}).

\subsection{Systematic Uncertainties}
\label{sec:Systematic}
In the previous section we quantified statistical uncertainties using the full covariance matrix obtained from 
the calibration of the FSUGold2R functional. However, such a purely statistical analysis from one model cannot 
assess systematic errors arising from biases and limitations intrinsic to the model. To estimate the linear dependence 
between the weak radii of calcium and argon, we consider an ensemble of 17 covariant energy density functionals;
the 16 functionals used in Ref.\,\cite{Reed:2021nqk} plus FSUGold2R. Although the correlation coefficient 
derived in this manner lacks a strict statistical interpretation, it remains an excellent indicator of linear 
dependence and a useful proxy for systematic trends.

Before proceeding further, we offer a word of caution. A genuine systematic study would include multiple 
families of energy-density functionals---both relativistic and nonrelativistic\,\cite{Piekarewicz:2012pp}---that 
reflect the intrinsic limitations of the theoretical framework that arise from the specific choice of functional,
biases, and  alternative fitting protocols. In the present paper, however, we restrict ourselves to the single 
class of covariant EDFs defined by the Lagrangian density introduced in the Formalism section. 

Motivated by the lack of experimental data to constrain the isovector sector, the set of 17 functionals was 
calibrated to span a broad range of values for the weak radius of ${}^{208}$Pb; see Fig.\ref{Fig1}. This 
spread provides a useful indicator of linear dependence between the weak radii of ${}^{208}$Pb and 
${}^{48}$Ca. Nevertheless, it is important to note that conclusions from such study may not capture 
systematics across different theoretical frameworks\,\cite{Piekarewicz:2012pp}.

\begin{figure}[ht]
\smallskip
 \includegraphics[width=0.75\columnwidth]{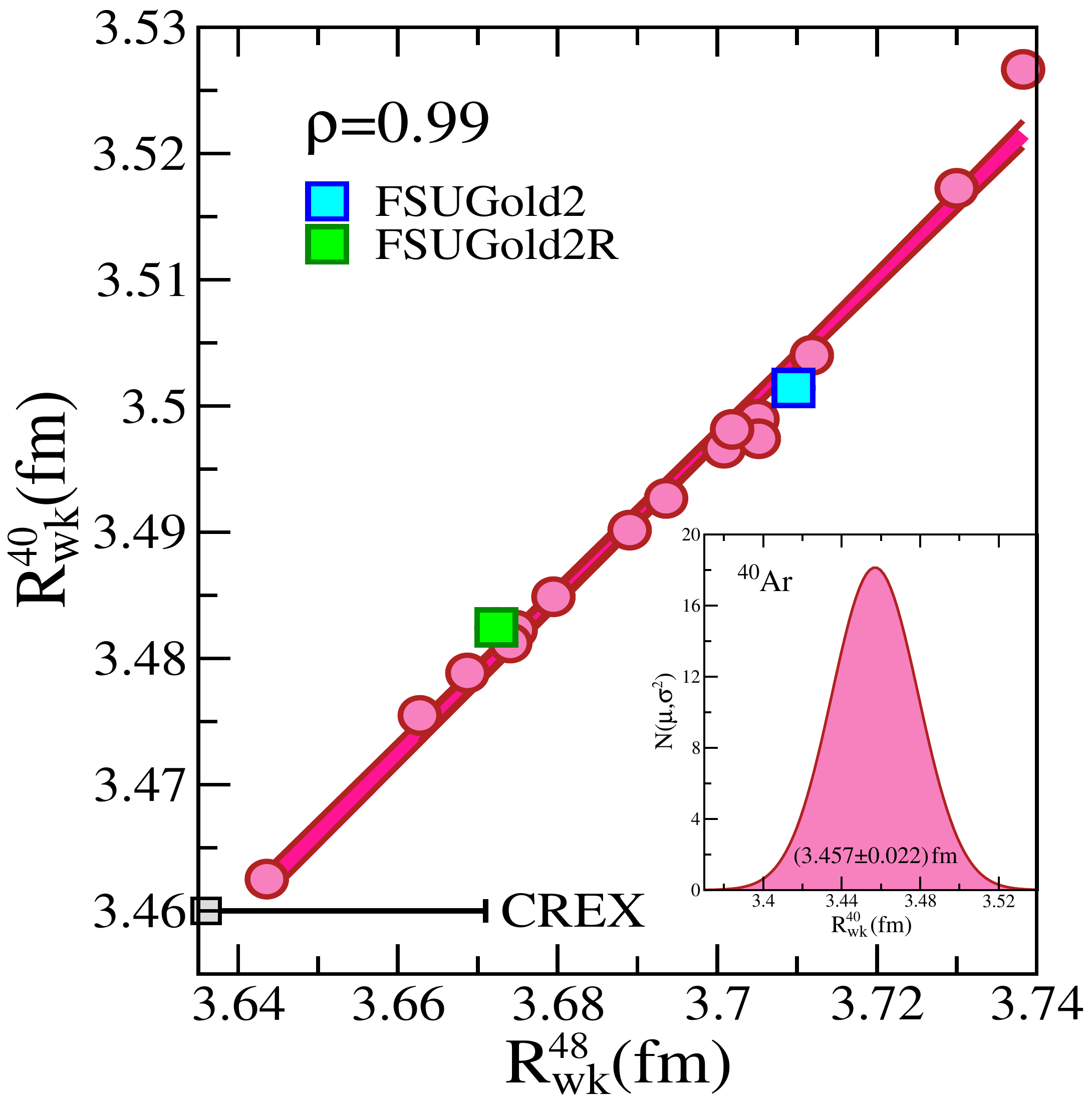}
\caption{(Color online) A  ``data-to-data" relation between the weak radii of ${}^{48}$Ca
               and ${}^{40}$Ar as predicted by the 17 covariant EDFs used in this work. The 
               inset displays the CREX-informed normal distribution inferred from the CREX
               results and the strong correlation between the two weak radii; see text for details.}
\label{Fig6}
\end{figure}

In Fig.\,\ref{Fig6} we show predictions from a set of 17 covariant EDFs for the weak radii of ${}^{48}$Ca 
and ${}^{40}$Ar. Before discussing the figure, we note that the small ($\sim\!0.3\%$) difference in the 
FSUGold2R central values of $R_{\rm wk}^{\,48}$ between Figs.\,\ref{Fig5} and~\ref{Fig6} arises from 
the treatment of the scalar-meson mass. In Fig.\,\ref{Fig6} the scalar mass is fixed to reproduce the 
charge radius of ${}^{208}$Pb\,\cite{Salinas:2023nci}, whereas in Fig.\,\ref{Fig5} it varies across the 
$10,000$ MCMC samples, yielding a slightly different average for $R_{\rm wk}^{\,48}$.

Returning to Fig.\,\ref{Fig6} and as noted earlier, the CREX central value lies below all model predictions, 
and only a few parametrizations fall within the CREX $1\sigma$ band. Nevertheless, the correlation 
coefficient reveals a strong linear dependence between the two observables. Mirroring the strategy 
implemented in the previous section, rather than relying on absolute model predictions, one employs 
this correlation to construct the $2\times2$ linear-regression covariance matrix. Applying 
Eqs.(\ref{LinCorr1})-(\ref{LinCorr2}) together with the CREX result yields the normal distribution shown 
in the inset of Fig.\,\ref{Fig6}. Based on systematic trends, one obtains the following CREX-informed 
estimate for the weak radius of ${}^{40}$Ar:
\begin{equation}
  R_{\rm wk}^{\,40} = (3.457 \pm 0.022)\,\text{fm}.
  \label{Rwk40Syst}
\end{equation}

Finally, using inverse-variance weighting of the statistical and systematic results provided
in Eqs.\,(\ref{Rwk40Stat}) and~(\ref{Rwk40Syst}), one obtains the following estimate for the 
weak radius of argon:
\begin{equation}
  R_{\rm wk}^{\,40} = 3.452 \pm 0.028~\text{(stat)} \pm 0.022~\text{(syst)}\,\text{fm}.
  \label{Rwk40Comb}
\end{equation}

\section{Conclusions}
\label{sec:Conclusions}

Coherent elastic neutrino–nucleus scattering (\Cevens) is a clean, purely electroweak process and a sensitive 
probe of new physics. The simplicity of the \Cevens cross section originates from its dependence (aside from 
simple kinematical factors) on just two Standard Model inputs: the Fermi constant and the  square of the mixing 
angle. Nevertheless, departures from full coherence---encoded in the weak nuclear form factor---can hinder the
identification of new physics.
Given the significant progress with liquid-argon detectors for \Cevens\!\!, dark-matter searches, and 
neutrino-oscillation experiments, placing stringent constraints on the weak radius of ${}^{40}$Ar---the 
one nuclear observable controlling the loss of coherence at small momentum transfers---was the main
goal of this paper.

Traditionally, constraints on the weak radius of ${}^{40}$Ar would have been obtained by sampling an ensemble 
of nuclear-structure models, each calibrated to a broad set of ground-state observables. In turn, each model 
would yield a prediction for $R_{\rm wk}^{40}$ with quantified statistical uncertainties. Such a collection of 
model enables an assessment of systematic errors. 

Unfortunately, the recent PREX and CREX results have called this strategy into question. State-of-the-art energy density 
functionals---both relativistic and nonrelativistic---fail to simultaneously reproduce the neutron-skin thicknesses of 
${}^{48}$Ca and ${}^{208}$Pb. Functionals with a stiff symmetry energy reproduce the ${}^{208}$Pb measurement 
but fail to do so for ${}^{48}$Ca, whereas models with a soft symmetry energy do the opposite.

To mitigate this tension, we reduced our reliance on absolute model predictions by invoking the CREX measurement. 
From a theoretical perspective, it was sufficient to show that FSUGold2R---a well-calibrated covariant EDF---as well as 
the collection of EDFs used in this work predict a strong linear correlation between the known weak radius of ${}^{48}$Ca 
and the unknown weak radius of ${}^{40}$Ar. Relying on this correlation and little else, the following CREX-informed 
estimate was obtained: $R_{\rm wk}^{\,40} = 3.452 \pm 0.028~\text{(stat)} \pm 0.022~\text{(syst)}\,\text{fm}$.

We close the paper with a few caveats to be considered in future work. First, all results presented here were 
obtained with a single class of covariant energy density functionals. To test any model dependence, one must 
verify that the correlations uncovered in this study are robust under a broader set of models, a step that will likely
broaden the systematic uncertainty. However, because the overall error is dominated by the CREX uncertainty, 
any increase in the systematic error may be mitigated by improving the current $\sim\!1\%$ CREX precision.
Second, unlike ${}^{48}$Ca and ${}^{208}$Pb, ${}^{40}$Ar is not doubly magic, so pairing correlations may be 
important given the small energy spacing between the $2s_{1/2}$-$1d_{3/2}$ proton orbitals. Simple estimates 
that vary the occupancies of these two orbitals induce changes in the weak radius of ${}^{40}$Ar of the order of 
$0.01\,\text{fm}$. Finally, electroweak spin-orbit currents are known to modify both charge and weak radii in nuclei 
with unpaired spin-orbit partners\,\cite{Horowitz:2012we}. In ${}^{48}$Ca, where the $\nu f_{7/2}$ orbital is filled 
and its $\nu f_{5/2}$ partner empty, spin-orbit currents increase $R_{\rm wk}^{\,48}$ by about $0.015\,\text{fm}$, 
comparable to the quoted uncertainties in Eq.(\ref{Rwk40Comb}). In the case of ${}^{40}$Ar with a smaller
neutron excess, the impact on the weak radius is roughly three times smaller, so spin-orbit corrections are 
expected to be subleading relative to the statistical and systematic errors quoted above.

In summary, with CREX as our main anchor and by relying on the robust calcium-argon correlation established here, 
we provide a benchmark for the weak radius of ${}^{40}$Ar that may be used as an experimental target for future 
\Cevens experiments with liquid argon and as a starting point for new-physics searches.

\bibliography{Main.bbl}

\begin{acknowledgments}\vspace{-10pt}
The author acknowledges many useful discussions with Pablo Giuliani, Caryn Palatchi, and Rex Tayloe.
This material is based upon work supported by the U.S. Department of Energy Office of Science, Office of 
Nuclear Physics under Award Number DE-FG02-92ER40750. 
\end{acknowledgments}

\vfill\eject
\end{document}